      [1999/12/01 v1.4c Il Nuovo Cimento]
\documentclass{cimento}
\usepackage{graphicx}  

\def\cl{\centerline}
\def\ni{\noindent}

\def\ms{\medskip}

\def\ea{et al.\,}
\def\eg{{\it e.g.\,}}
\def\be{\begin{equation}}
\def\ee{\end{equation}}
\def\apj{ApJ}
\def\apjl{ApJL}
\def\aa{A\&A}
\def\mn{MNRAS}
\def\prd{Phys. Rev. D}
\def\rel{relativistic\,\,}
\def\nrel{nonrelativistic\,\,}

\title{The Sunyaev-Zeldovich Effect}
\author{Y.~Rephaeli\from{ins:x}\from{ins:y}\ETC,
S.~Sadeh\from{ins:x}
        \atque
M.~Shimon\from{ins:x}\from{ins:y}}
\instlist{\inst{ins:x} School of Physics \& Astronomy, Tel Aviv 
University, Tel Aviv, Israel\\
\inst{ins:y}Center for Astrophysics and Space Sciences, UCSD, La Jolla, CA, USA}

\begin{document}

\maketitle

\section{Introduction}

During passage through a cluster of galaxies some of the photons 
of the cosmic microwave background (CMB) radiation are scattered by 
electrons in the hot intracluster (IC) gas. The scattering slightly 
modifies the incident Planck spectrum, imprinting on it a unique 
spectral signature that was first described by Sunyaev \& Zeldovich 
(1972). [The impact of hypothetical hot intergalactic gas on the 
CMB spectrum was calculated earlier by Weymann (1966) and Zeldovich 
\& Sunyaev (1969).] This Sunyaev-Zeldovich (S-Z) effect is a valuable 
tool for probing the cluster environment and the global properties 
of the universe. The basic significance of the effect 
was pointed out in the early work of Sunyaev \& Zeldovich, and in 
many papers written during the first decade following their original 
paper. General reviews of the effect and its measurements include 
those by Rephaeli (1995a), Birkinshaw (1999), and Carlstrom \ea (2002).

The major challenge of measuring the effect in nearby clusters was 
taken up soon after the effect was identified, but some decade and 
a half passed before convincing detections were made. Observational 
results from single-dish radio measurements were reviewed  by 
Birkinshaw (1999). Growing realization of the cosmological 
significance of the effect has led to major improvements in 
observational techniques, and to extensive theoretical investigations 
of its many aspects. The use of interferometric arrays and the 
substantial progress in the development of sensitive radio receivers 
led to first interferometric images of the effect (Jones \ea 1993, 
Carlstrom \ea 1996). Some 60 clusters have already been imaged with 
the OVRO and BIMA arrays (Carlstrom \ea 2002).

With the many S-Z projects that will become operational in the 
very near future, measurements of the S-Z effect are about to expand 
tremendously. The effect is expected to be detected in thousands of 
clusters when planned cluster surveys are conducted. For many clusters 
we expect to have detailed spatial mapping of the effect. Together with 
the much expanded spectral coverage, a large S-Z database will greatly 
advance our ability to fully exploit the potential of using the effect 
as a {\it precise} cosmological probe. The main observational challenges 
in precision S-Z work are discussed in the review of Birkinshaw in this 
volume. As noted, general reviews are available of the S-Z effect and 
the main observational results. Our brief review here is meant to be 
somewhat more pedagogical.

\section{The Effect}

The spectral change resulting from Compton scattering of the CMB 
by IC gas was calculated by Sunyaev \& Zeldovich (1972) in the 
non-relativistic limit based on a solution to the Kompaneets equation 
(which essentially is a diffusion approximation to the exact kinetic 
equation). As was clearly demonstrated by Rephaeli (1995b), the 
approximate (though elegantly simple) expressions for the intensity 
change, $\Delta I$, obtained by Sunyaev \& Zeldovich (eqs. 6-9) are 
not sufficiently accurate at high frequencies and temperatures. IC 
gas temperatures span the range $3 - 15$ keV; high electron velocities 
result in relatively large photon energy change in the scattering, 
requiring a more exact \rel calculation.

\subsection{\bf Total Intensity Change}

Using the exact probability distribution, and a relativistically
correct form of the electron Maxwellian velocity distribution,
Rephaeli (1995b) calculated the resulting intensity change in the 
limit of small optical depth to Thomson scattering, $\tau$, keeping 
terms linear in $\tau$. In this semi-analytic treatment for the 
calculation of $\Delta I_t$ -- the change of intensity due to 
scattering by electrons with a thermal velocity distribution -- the 
starting point is the probability of scattering of an incoming photon 
(direction $\mu_{0}=\cos\theta_{0}$) to the direction 
$\mu'_{0}=\cos\theta'_{0}$ is (Chandrasekhar 1950)
\begin{eqnarray}
f\left(\mu_{0},\mu'_{0}\right)=\frac{3}{8}\left[1+
\mu_{0}^{2}\mu'^{2}_{0}+\frac{1}{2}\left(1-\mu_{0}^{2}\right)
\left(1-\mu'^{2}_{0}\right)\right]
\label{fmumu'}
\end{eqnarray}
where the subscript 0 refers to the electron rest frame. The resulting 
frequency shift is
\begin{eqnarray}
s=\ln\left(\nu'/\nu\right)=\ln\left(\frac{1+\beta\mu'_{0}}{1+
\beta\mu_{0}}\right)
\label{s}
\end{eqnarray}
where $\nu$, $\nu'$ are the photon frequency before and after the 
scattering, and $\beta=v/c$ is the dimensionless electron velocity 
in the CMB frame. It is somewhat more convenient to use the variables 
$\mu,s$ and $\beta$ instead of $\mu,\mu'$ and $\beta$. The probability 
that a scattering results in a frequency shift $s$ is (Wright 1979)
\begin{eqnarray}
\mathcal{P}\left(s,\beta\right)=
\frac{1}{2\gamma^{4}\beta}\int\frac{e^{s}f\left(\mu_{0},
\mu'_{0}\right)}{\left(1+\beta\mu_{0}\right)^{2}}d\mu_{0}.
\label{P}
\end{eqnarray}
Averaging over a Maxwellian distribution for the electrons yields
\begin{eqnarray}
\mathcal{P}_{1}\left(s\right)=\frac{\int\beta^{2}\gamma^{5}e^{-
\frac{\left(\gamma-1\right)}{\Theta}}\mathcal{P}\left(s,
\beta\right)d\beta}{\int\beta^{2}\gamma^{5}e^{-\frac{\left(\gamma-
1\right)}{\Theta}}d\beta} , 
\label{P1}
\end{eqnarray}
where $\Theta= kT_{e}/m_{e}c^{2}$. The total change in the photon 
occupation number along a line of sight (los) to the cluster can 
now be written as
\begin{eqnarray}
\Delta n_{t}(x)=\tau\int_{-\infty}^{\infty}\left[n\left(xe^{s}\right)-
n\left(x\right)\right]\mathcal{P}_{1}\left(s\right)ds
\label{deltant}
\end{eqnarray}
where $x$ is the dimensionless frequency, $x=h\nu/kT$, $T$ is the CMB 
temperature, the optical depth is $\tau\equiv\sigma_{T}\int n_{e}dl$, 
and $\sigma_T$ is the Thomson cross section. The change of intensity 
is simply calculated using $\Delta I_{t}(x)=i_{o}x^{3}\Delta n_{t}(x)$, 
where $i_{o}=2(kT)^3 /(hc)^2$.

In the \nrel limit the Kompaneets eq. can be readily solved leading to 
a simple analytic expression (Sunyaev \& Zeldovich 1972) for 
$\Delta I_{t}$,
\be
\Delta I_{t} = i_{o} y g(x) \;,
\ee
where the dependence on the cluster gas density ($n$) and temperature 
is in the Comptonization parameter (essentially, an integral over the 
electron pressure),
\be
y=\int(kT_e/mc^2) n \sigma_T dl \:.
\ee
The spectral function 
\be
g(x)={x^4e^x \over (e^x -1)^2} \left[{x (e^x +1)\over e^x -1}-4
\right],
\ee
is negative for $x < 3.83$ and positive at larger values of this 
crossover frequency,  $\sim 217$ GHz. The magnitude of the relative 
temperature change due to the thermal effect is $\Delta T_t/T = -2y$ in 
the R-J region, with $y \sim 10^{-4}$ along a line of sight (los) through 
the center of a rich cluster.

The motion of the cluster in the CMB frame induces a {\it kinematic} 
(Doppler) S-Z component, which can be easily calculated in the \nrel limit 
(Sunyaev \& Zeldovich 1980)
\be
\Delta I_k = - i_{o} \tau \beta_{c} h(x)  \;, \;\;\;\; 
h(x) = {x^{4}e^{x}\over (e^x -1)^2} \; ,
\ee
where $\beta_{c} = v_r/c$, with $v_r$ the line of sight component of the 
cluster peculiar velocity. The corresponding temperature change for this 
kinematic component is $\Delta T_k/T= -\tau \beta_{c}$.

\begin{figure}[t]
\cl{\includegraphics[height=4in,width=4in]{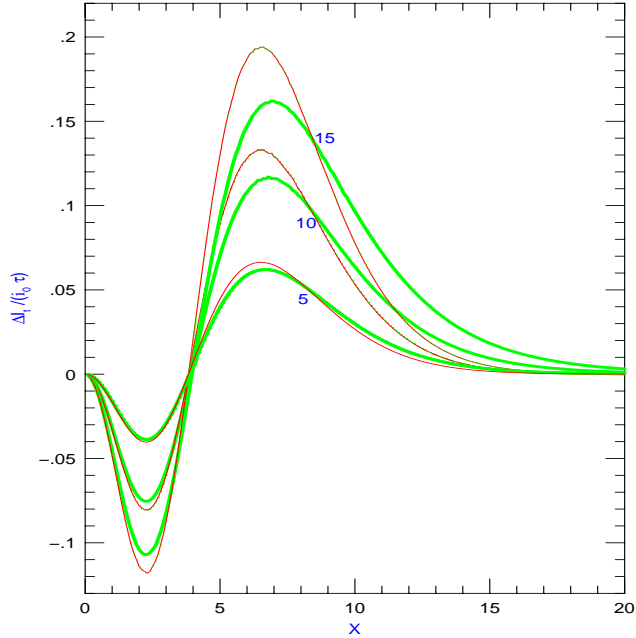}}
\caption{The spectral distribution of $\Delta I_t /(i_{o} \tau)$. 
The pairs of thick (green) and thin (red) lines, labeled with 
$kT_e =$ 5, 10, and 15 keV, show the \rel and nonrelativistic 
distributions, respectively.}
\end{figure}  

The results of the \rel and \nrel calculations of $\Delta I_{t}(x)$ are 
shown in fig. 1 for $kT_e =5, 10, 15$ keV. As is evident, the more 
accurate \rel calculation yields values that are appreciably different 
from those based on the \nrel formula. The deviations increase with 
$T_e$ and are particularly large near the crossover frequency, where the 
purely thermal effect vanishes. This frequency shifts to higher values 
with increasing $T_e$. Also, while the dependence $\Delta I_{t}$ 
on $\tau$ is still linear to a high degree of accuracy, its dependence 
on $T_e$ is sufficiently non-linear (for typical values of $T_e$) that 
the intensity change should no longer be determined through the use 
of the Comptonization parameter $y$.

While most S-Z observations have traditionally been at
(relatively) low frequencies ($\sim 30$ GHz, or $x \sim 0.5$) where
the approximate \nrel description is roughly adequate, some  
16 clusters were observed also at much higher frequencies 
(with, \eg, the SuZIE and MITO telescopes, up to $x \sim 6.2$). 
In fact, most future S-Z projects will observe the effect at 
several high frequencies. Use of the relativistically exact expressions
for $\Delta I_t$ and $\Delta I_k$ is clearly necessary at $\nu >> 30$ GHz, 
especially when the effect is used to determine precise values of
cluster and cosmological parameters. Moreover, since the ability
to measure peculiar velocities of clusters depends very much on 
observations very close to the crossover frequency, its dependence 
on $T_e$, which is approximately given (Shimon \& Rephaeli 2004) in 
\be
x_0 \simeq 3.8300(1+1.1206\theta + 2.0783\theta^{2}-80.7481\theta^{3}) ,
\ee
should be taken into account. Note also that because high precision 
S-Z work entails use of X-ray derived gas parameters, similarly 
accurate expressions for the X-ray bremsstrahlung emissivity have to 
be employed (Rephaeli \& Yankovitch 1997). In the latter paper first 
order relativistic corrections to the electron velocity distribution 
and electron-electron bremsstrahlung were taken into account in 
correcting values of the Hubble constant, $H_0$, that were previously 
derived using the \nrel expression for the emissivity (see also Hughes
\& Birkinshaw 1998, and Nozawa \ea 1998b).

The calculation of Rephaeli (1995b) motivated various generalizations and 
extensions of the relativistic treatment. Challinor \& Lasenby (1998) 
obtained an analytic approximation to the solution of the relativistically 
generalized Kompaneets equation, whose accuracy was then extended to 
fifth order in $\Theta$ (Nozawa \ea 1998a). Sazonov \& Sunyaev (1998) 
and Nozawa et al. (1998b) have extended the relativistic treatment 
also to the kinematic component obtaining (for the first time) the 
leading cross terms in the expression for the combined (thermal and 
kinematic) intensity change, $\Delta I_t + \Delta I_k$, which depend 
on both $T_e$ and $v_{r}$. Since in some rich clusters 
$\tau \sim 0.02-0.03$, sufficiently accurate analytic expansions 
of the expression for $\Delta I$ in powers of $\Theta = kT_{e}/mc^2$ 
necessitate the (consistent) inclusion also of multiple scatterings 
of order $\tau^2$ (Molnar \& Birkinshaw 1999, Itoh \ea 2000, Shimon 
\& Rephaeli 2004). Details on the calculation of the full effect 
are given in these papers. 

The exact \rel calculation of $\Delta I$ does not lead to a simple 
analytic expression. In order to obtain an approximate analytic 
expression (that can greatly simplify the analysis of S-Z measurements), 
one needs to expand the formal expression for $\Delta I$ in powers of 
(the small quantities) $\tau$, $\Theta$, and $\beta_{c}$. For the 
resulting expression to be accurate to within to $\sim 2\%$ for $kT_{e} 
< 50$ keV, Shimon \& Rephaeli (2004) included terms up to 
$\tau \Theta^{12}$, $\tau^{2} \Theta^{5}$, and $\beta_{c}^{2}\Theta^{4}$. 
[Itoh \& Nozawa (2004) have derived a slightly more accurate fit to the 
results of the exact numerical calculation; however, results of this fit 
are given in terms of an array of tabulated coefficients.] Since cluster 
velocities are expected to be generally well below $1000$ km/s, terms 
quadratic in $\beta_{c}$ can be ignored. Doing so we can write the total 
intensity change as the sum 
\be
\Delta I/i_{o} = \tau \sum_{j=1}^{8}f_{j}(x)\Theta^{j} 
+ \tau^{2} \sum_{j=1}^{4} f_{j+8}(x)\Theta^{j+1} - \tau \beta_{r} 
\left[h_{0}(x) + \sum_{j=1}^{4} f_{j+12}(x) \Theta^{j} \right] ,
\ee
where $f_{j}(x)= x^{3}F_{j}(x)$, with $F_{j}(x)$ defined in Shimon \&
Rephaeli (2004).

\subsection{\bf Polarization Components}

Due to the particular angular dependence of the cross section, incident 
unpolarized radiation is linearly polarized if it has a finite 
quadrupole moment. The polarization - which is described in a plane 
orthogonal to the los - is completely specified by the $Q$ and $U$ 
Stokes parameters. Axial symmetry of the scattering (as is evident 
from the dependence of the frequency change, Equation 2, on 
$\theta$, $\theta'$ but not on $\phi$ or $\phi'$) allows the freedom 
to select a frame such that $U=0$. [Note that this is no longer the 
case in the presence of a magnetic field (\eg Ohno et al. 2003) or 
when the cluster substructure is spatially resolved (Shimon \ea 2005).] 
Using the above notation $Q$ can be written as 
\be
Q(\mu)=\frac{3}{8}\tau(1-\mu_{0}^{2})\int_{-1}^{1}P_{2}(\mu_{0}')
I(\mu_{0}')d\mu_{0}' ,
\ee
where $P_{2}(\mu'_{0})$ is the second Legendre polynomial. From the 
orthogonality of the Legendre polynomials it is clear that $Q$ depends 
only on the quadrupole moment of the incident radiation as observed in 
the electron rest frame. 

CMB polarization resulting from Compton scattering in clusters was 
first considered by Sunyaev \& Zeldovich (1980). Using Equation (12) 
they obtained, to first order in $\tau$, 
\begin{eqnarray}
Q_{1}=\frac{\tau}{20}\beta_{t}^{2}\frac{xe^{x}(e^{x}+1)}
{(e^{x}-1)} ,
\end{eqnarray}
and to second order
\begin{eqnarray}
Q_{2}=\frac{\tau^{2}}{40}\beta_{t}. 
\end{eqnarray}
Both components, which are given here in (temperature) K, depend 
only on the tangential velocity of the cluster, $c\beta_{t}$, and 
thus their measurement provides complementary velocity information 
to (the leading term in) the kinematic S-Z effect, which depends on 
the los velocity.

\begin{figure}[h]
\cl{\includegraphics[height=4in,width=4in]{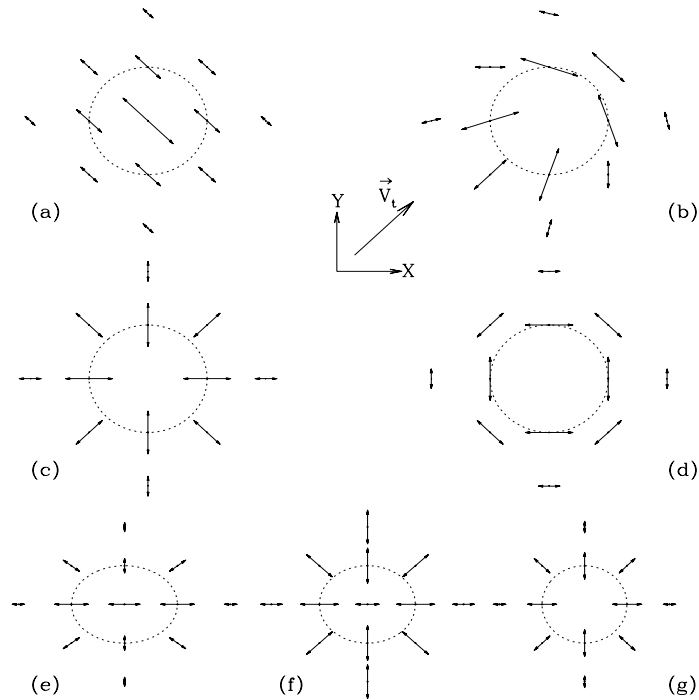}}
\caption{Patterns of the polarization components described in the 
text. In cases (a)--(d) $\beta$-profile with $\beta=3/2$ is assumed, 
and the cluster peculiar velocity is as indicated by the arrow 
$\vec{V}_{t}$. (a) The $\propto\beta_{t}^2\tau$ component. (b) The
$\propto\beta_{t}\tau^2$ component. (c) The thermal component
$\propto(kT_e/m_ec^2)\tau^2$ at frequencies $x<3.83$, and (d) at 
$x> 3.83$. Elliptical cluster with the ratio of the principal 
axes $b/a=0.8$ and the symmetry (longer) axis in the $X$ direction 
(e), the symmetry axis inclined at $45^\circ$ to the picture plane 
(f), and with the symmetry axis along the line of sight (g). Figure 
adopted from Sazonov \& Sunyaev (1999).}
\end{figure}

The proportionality of $Q_{1}$ to $\beta_{t}^{2}$ follows from its 
dependence on the second derivative of the Planck distribution, while 
$Q_{2}$ is proportional to its first derivative. This feature renders 
these two effects distinguishable through multi-frequency observations. 
Moreover, due to the different dependence on the optical depth, the 
two effects are expected to be distinguishable by their different 
spatial signatures in spherical clusters; these are shown in fig. 2a 
\& 2b), along with some of the other polarization components that 
were calculated and described by Sazonov \& Sunyaev (1999) and mentioned 
briefly below. Corrections to the $\propto\tau\beta^{2}$ component 
(Equation 13) due to the random (thermal) component of electron velocity 
were calculated by Challinor, Ford \& Lasenby (2000), and Itoh et al. 
(2000). These corrections generally amount to $\sim 10\%$ reduction 
in $Q_{1}$. 

Another polarization component is due to multiple scatterings; the 
first scattering induces the requisite quadrupole moment; the radiation 
is polarized upon second scattering. This component is proportional to 
$\tau^{2}\Theta$ (figs. 2c \& 2d).

In addition to the above polarization components due to quadrupole moments 
associated with electron motions (both directed and random), polarization 
arises also from the intrinsic CMB quadrupole of the primary anisotropy. 
This achromatic component is $\frac{\tau}{10}\sqrt{\frac{15}{2\pi}C_{2}}$, 
where $\sqrt{C_{2}}$ is the quadrupole moment of the CMB power spectrum, 
$\sqrt{C_{2}}\sim 1.9\times 10^{-5}$). The magnitude of this component 
is typically $\sim 20$ nK; although small, it could possibly be 
distinguished from the other components by its random distribution over 
the sky. It was proposed that the dependence of this polarization component 
on $C_{2}$ can be used to reduce cosmic variance (Kamionkowski \& Loeb 1997), 
and that the evolution of $C_{2}$ with redshift provides a mean to probe 
dark energy models (Cooray \& Baumann 2003).
 
In the calculations of the above polarization components it is usually 
assumed that the cluster is spherically symmetric. If the cluster is 
ellipsoidal the optical depth is anisotropic and the polarization 
patterns are as shown in figs. 2e-2g. 

Measurement of polarized S-Z signals has been a major challenge so far. 
Based on the expected high sensitivity and high spatial resolution of 
upcoming polarization experiments, the detection of the major signals 
is projected to be feasible. This has motivated more detailed studies 
of CMB polarization taking into consideration the complex morphology of 
evolving clusters that can be followed in hydrodynamical simulations. 
For example, Diego, Mazzotta \& Silk (2003) and  Lavaux \ea (2004) 
explored the polarization induced by bulk motions during merging of 
subclusters. The fact that both the magnitudes and spatial patterns 
of polarized cluster signals cannot be realistically explored in the 
context of idealized gas density and temperature distributions was 
made clear by the work of Shimon \ea (2005). In their paper the leading 
polarization components ($\propto\tau^{2}\Theta$, $\propto\tau\beta^{2}$ 
and $\propto\tau^{2}\beta$) were calculated in a rich cluster simulated 
by the Enzo code. The maximal polarization levels were found to be a few 
tenths of a $\mu$K, with patterns that are considerably more complex than 
predicted by the idealized calculations whose results are shown in 
fig. 2. 

In addition to polarization in individual clusters the impact of the 
cluster population on the CMB polarization is of considerable interest. 
This has been studied by, \eg Hu (2000), and is discussed by Cooray, 
Baumann \& Sigurdson in this volume.

\section{Power Spectrum \& Cluster Counts}

It has been realized early on that the scattering of the CMB in clusters
induces spatial anisotropy (Sunyaev 1977) whose basic properties were
already determined in the context of a simple model for the distribution
of clusters and the evolution of IC gas (Rephaeli 1981). More than two
decades later it is now well established that this is the most important
secondary anisotropy on arcminute scales. Since the anisotropy arises
from scattering of the CMB in the evolving population of clusters,
its power spectrum and cluster number counts can potentially yield
important information on the properties of IC gas, the cluster mass   
function, cosmological evolution of clusters and their gaseous contents,
as well as some of the global cosmological and large scale structure
parameters. Clearly, therefore, the detailed description of this
anisotropy necessitates also the additional modeling of gas properties
across the evolving population of clusters.

The traditional approach to the calculation of the S-Z anisotropy is
based on the Press-Schechter model for the cluster mass function,
$n(M,z)$, the comoving density of clusters of mass M at redshift z.
Following collapse and virialization, IC gas is presumed to have reached
hydrostatic equilibrium at the virial temperature, with a density
distribution that is commonly assumed to have an isothermal $\beta$
profile. The mass function is normalized by specifying the mass variance
on a scale of $8h^{-1}$ Mpc, $\sigma_8$, a parameter which can be
determined from the primary CMB power spectrum, from large-scale
galaxy surveys, or from the observed X-ray temperature function when 
calibrated by a mass-temperature relation (which is currently limited 
to clusters at relatively low redshifts). The cluster-induced anisotropy 
has been studied at an increasingly greater degree of sophistication 
and detail (and in a range of cosmological and dark matter models) 
beginning more than a decade ago (\eg, Makino \& Suto 1993, Bartlett \& 
Silk 1994, Colafrancesco \ea 1994). For example, in the latter paper the
temperature anisotropy was calculated in a flat CDM model including gas
evolution. The approach adopted in that paper was later (Colafrancesco
\ea 1997) extended to other cosmological models, and to the calculation
of the mass and redshift distributions of the many thousands of
clusters that are expected to be detected during the planned Planck
survey. The anisotropy and its power spectra can also be generated
directly from hydrodynamical simulations (\eg, da Silva \ea 2000).
And, of course, the range of cosmological models was extended to
include currently favored $\Lambda$CDM models (beginning with the
works of Komatsu \& Kitayama 1999, Molnar \& Birkinshaw 2000, and
Cooray \ea 2000).

The main features of the power spectra of the anisotropy due to
the thermal and kinematic S-Z components, and the cluster number
counts, were determined already in the above mentioned papers.   
Given the nature of the required input parameters characterizing   
the cosmological and cluster models, it is not surprising that the
predicted power spectra and number counts span a wide range.
We briefly review some of the recent (since 2001) results from
analytical calculations and hydrodynamical simulations in $\Lambda$CDM  
models.

Two approaches have been adopted in the calculation of the S-Z induced 
anisotropy. In the analytic approach clusters are described by simple 
models, which characterize their morphologies, temperatures, and their 
gas content and evolution, usually as functions of the virial mass and 
redshift. For each cluster the profile of the Comptonization parameter 
is calculated and transformed into Fourier space. The overall power 
spectrum can then be quantified by convolving the transformed 
$y$-parameter with a cosmological mass function characterizing
the universal cluster population, traditionally taken to be of the 
Press \& Schechter (1974) form. The power spectrum can then be  
calculated numerically using the expression 
\begin{equation}
C_{\ell}=\int_{z} r^{2}\,\frac{dr}{dz}\int_{M}
N(M,z)\,\zeta_{\ell}(M,z)\,dM\,dz,
\end{equation}
where $r$ is the comoving radial distance to a cluster of mass $M$
located at redshift $z$, $N(M,z)$ is the mass function, and 
$\zeta_{\ell}(M,z)$ is the angular Fourier transform of the $\Delta T$.
In the second approach high resolution dynamical and hydrodynamical 
simulations (incorporating both dark matter and baryonic components) 
yield large cosmological fields which enable the identification of 
individual clusters and their properties relevant to the calculation 
of $y$, and the S-Z power spectrum. 

Of course, these approaches have their respective advantages and 
drawbacks. Simple analytic modeling in the first approach can be 
readily implemented in a computer code, and the impact of various 
input data can be easily assessed. The main disadvantage of this 
approach is the explicit need for detailed modeling of each component 
of the calculation, each with its own uncertainties due to either 
oversimplified physics, or lack of quality data, or both. An example 
for this is the mass-temperature relation, used to attribute a 
temperature to a cluster with virial mass $M$. A favored scaling is 
the virial relation 
\begin{equation}
T=T_0\,(1+z)\left(\frac{M}{10^{15}\,h^{-1}\,M_{\odot}}\right)^{2/3}\,
\Omega_0^{1/3}\,\left[\frac{\Delta(\Omega_0,z)}
{\Delta(\Omega_0=1,z=0}\right]^{1/3},
\end{equation}
where $T_0$ is the gas temperature of a $10^{15}\,h^{-1} \, M_{\odot}$ 
cluster located at redshift $z=0$, and $\Delta(\Omega_0,z)$ is the 
non-linear density contrast at redshift $z$, and $h$ is the value of 
$H_0$ in units of $100$ km s$^{-1}$ Mpc$^{-1}$. In addition to the 
questionable assumption that clusters are relaxed and in hydrostatic 
equilibrium, there is appreciable uncertainty both in the value of 
$T_0$ and the scaling of the temperature with redshift. The main 
current limitations of cluster hydrodynamical simulations are the 
inclusion of a restricted range of physical processes, and 
insufficient spatial resolution.

As consequence of the redshift independence of the S-Z effect the    
detection of many distant clusters by upcoming projects is very 
feasible. Obviously, the number of clusters that can in principle be 
detected depends on both their universal population, internal properties, 
and the flux detection limit of the experiment. The number of clusters 
whose S-Z flux exceeds $\Delta F_{\nu}$ is
\begin{equation}
N(>\Delta F_{\nu})=\int r^{2}\frac{dr}{dz}dz
\int_{\Delta F_{\nu}}
B(M,z)N(M,z)\,dM .
\end{equation}
The lower limit of the mass integral corresponds to the limiting flux 
from a cluster with mass $M$ located at redshift $z$, and $B(M,z)$ is 
either unity or zero, depending on whether the flux measured from the 
cluster is higher or lower than the flux limit of the experiment. 
Since the detection of clusters through the S-Z effect cannot reveal 
any information concerning their redshifts, the observable quantity is 
the cumulative number clusters at all redshifts. Note that just as for 
the power spectrum, theoretical predictions of cluster number counts 
span a wide range that reflects the uncertainties in internal properties 
of clusters and their mass function.

The power spectrum of the S-Z effect in a $\Lambda$CDM model has  
been calculated by numerous authors (e.g. Komatsu \& Kitayama 1999,
Molnar \& Birkinshaw 2000, Cooray \ea 2000, Komatsu \& Seljak 2002, 
Bond \ea 2002, Zhang \ea 2002, Springel \ea 2001) both analytically 
and from results of simulations. The resulting power spectra have 
essentially a universal shape - a steep monotonic ascent to a peak 
followed by a sharp descent, as shown in the (semi-logarithmic) plot 
in fig. 4. The curves differ, however, in the multipoles at which 
the peaks fall, the magnitude of the peak power, and - to a lesser 
extent - the width of the peaks, as detailed in the rest of this section.

The power spectrum was calculated by Molnar \& Birkinshaw (1999) for 
an $\Omega_0=0.2$, $\Omega_{\Lambda}=0.8$, $h=0.5$ model, using a Press 
\& Schechter mass function normalized by $\sigma_8=1.35$, a density 
fluctuation spectrum with index $n=-1.4$, and cluster mass range bounded 
by $M_{min}=10^{13}\,h^{-1}\,M_{\odot}$ and $M_{max}=
10^{16}\,h^{-1}\,M_{\odot}$. Evolving gas fraction was adopted, with 
the density decreasing according to $\sim (1+z)^{-1}$ at earlier times. 
An isothermal gas was assumed, with a temperature of $8.7\,keV$ 
ascribed to a local cluster of mass $10^{15}\,M_{\odot}$, and a $\beta$ 
density profile with $\beta=1$. The resulting power spectrum peaks at 
$\ell\sim 2000$, with a peak magnitude of $\ell(\ell+1)/2\pi\cdot 
C_{\ell}\sim 10^{-11}$ K$^2$. Komatsu \& Kitayama (1999) employed
$\Omega_{\Lambda}=0.5$, $\Omega_0=0.3$, and $h=0.7$, similarly modeled 
the gas profile and mass function, but adopted $\sigma_8=1$, $\beta=2/3$,
$M_{min}=5\cdot 10^{13}\,h^{-1}\,M_{\odot}$, and $M_{max}=5\cdot 
10^{15}\,h^{-1}\,M_{\odot}$. Moreover, the gas content was assumed to 
be non-evolving, and the core radius was determined by assuming that 
the gas is in a minimum entropy phase, which implies an evolving core 
size. This model was formulated using a free parameter, $\epsilon$, 
with the values $-1,0,1$. The power spectrum peaks at $l\sim 2000-4000$, 
with maximal power in the range $4-7\cdot 10^{-12}\,K^2$ for the three 
values of $\epsilon$. Note that the spectral function $g(x)$ was 
divided out, so that in order to compare these results with others 
calculated in the Rayleigh-Jeans region, where $g(x)=-2$, the power 
spectrum magnitude must be multiplied by a factor of $4$. Cooray \ea 
(2000) calculated the power spectrum for an $\Omega_0=0.65$,
$\Omega_{\Lambda}=0.65$ and $h=0.65$ model, normalized the Press 
\& Schechter mass function with $\sigma_8=0.9$, and determined the 
gas profile from a solution to the hydrostatic equation with an 
NFW-distributed dark matter profile. The gas was assumed to be 
non-evolving fraction of the total mass, and isothermal with a (local) 
$10^{15}\,h^{-1}\,M_{\odot}$ cluster assigned to have a temperature 
of $5.2\,keV$. The resulting peak power is $\sim 10^{-11}\,K^2$ at 
$\ell\sim 2000-3000$.

Studies of the S-Z power spectrum based on hydrodynamical simulations 
include those of Springel \ea (2001), and Bond \ea (2002), who adopted 
the favored $\Omega_{\Lambda}=0.7$, $\Omega_0=0.3$, model with 
$\sigma_8=0.9$ and index $n=1$. The resulting power spectra are  
broad and peak at considerably higher multipoles of up to 
$\sim 10000$, with peak power that is also somewhat higher than 
calculated analytically. (Similar results were obtained also by 
Zhang \ea 2002.)

As an example of how different models may lead to different power
predictions we show in fig. 3 a comparison of power spectra 
calculated with three different mass functions, Press \& Schechter 
(1974), Lee \& Shandarin (1999), Sheth \& Tormen (1999), hereafter 
PS, LS and ST, respectively.
\begin{figure}[h]
\centering
\includegraphics[height=3in,width=3in]{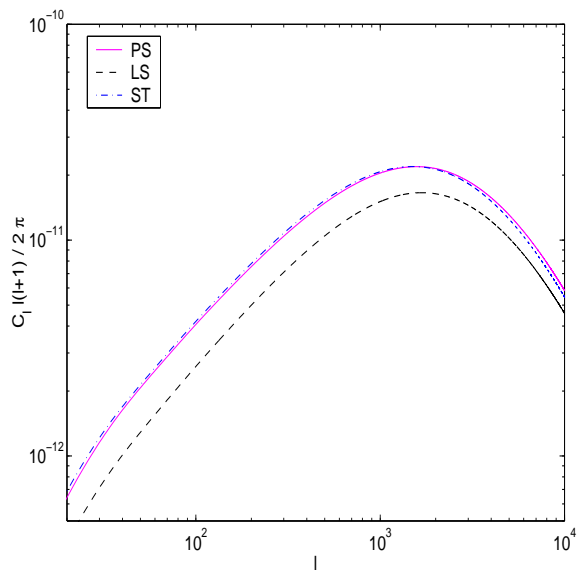}
\caption{Dependence of the S-Z power spectrum on the choice of mass
function.}
\end{figure}
While the PS and ST mass functions seem to yield approximately the 
same results, the LS mass function generates lower magnitude power at
all multipoles by $\sim 30-50\%$.

From these and numerous other published results it is clear that 
the predicted power spectra span a wide range of values for the 
width of the peak and its magnitude, reflecting mostly (but not 
only) different assumed values $\sigma_8=0.9$ and gas properties. 
The impact of a particular choice of parameters on the peak power 
and its typical scale ($\ell$) can be readily predicted. Well 
known is the steep dependence of the power on $\sigma_8$ (fig. 4), 
due to its location in the exponential part of the mass function; 
various studies have shown that even a mild change of $\sim 10\%$ 
may lead to an order of magnitude difference in the magnitude of 
the power spectrum. On the other hand, the spectrum is much less 
sensitive to other cosmological parameters, such as $\Omega_m$, 
implying that the uncertainty in $\sigma_8$ is of major relevance 
to precise determination of the power spectrum. The impact of 
different choices of IC gas properties can be easily anticipated; 
for example, an increase of either the normalization of the 
mass-temperature relation or the central electron density by a 
factor $f$, would lead to an increase of power by a factor $f^{2}$, 
since the power spectrum is proportional to $y^2$. A steeper $y$ 
profile - as is obtained in polytropic gas models - shifts power 
towards higher multipoles since the effective size of the cluster 
decreases (Sadeh \& Rephaeli 2004). This is also the case if the 
gas is assumed not to evolve with redshift, since this leads to 
an increase in the population of distant, smaller clusters. 

\begin{figure}[h]
\centering
\includegraphics[height=3in,width=3in]{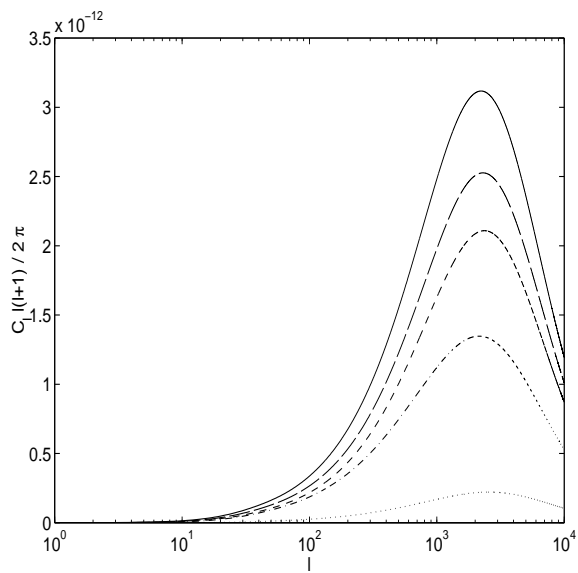}
\caption{Power spectrum predictions for various $\sigma_8$
  values. Solid, long dashed, short dashed, dashed-dotted and dotted
  curves correspond to $\sigma_8=1.32,1.23,1.18,1.00,0.68$, respectively.}
\end{figure}

Along with calculations of the power spectrum, many studies have been 
conducted of cluster number counts; the detailed predictions for these 
too show a large variance. Colafrancesco \ea (1997) used a 
$\Lambda$CDM model ($\Omega_0=0.2, \Omega_{\Lambda}=0.8, h=0.5, n=1$)
and four different flux limits ($10-60\,mJy$) to assess the distribution 
of cumulative number counts as a function of redshift. Their 
calculations were carried out within the framework of both evolving 
and non-evolving IC gas scenarios; for a flux limit of $30\,mJy$ they 
found cumulative counts of $\sim 2\cdot 10^4$ clusters to a redshift of 
$z\sim 0.1$ in the non-evolving gas scenario, and $\sim 4\cdot 10^{3}$ 
clusters when IC evolution is taken into account. In both models counts 
drop drastically around $z\sim 1$, but with the number of clusters at 
higher redshifts still much larger in the non-evolutionary case. 
Number counts exhibit the same degree of sensitivity to $\sigma_8$ as
the power spectrum; a graphical demonstration of this effect is shown
in fig. 5.
\begin{figure}[h]
\centering
\includegraphics[height=3in,width=3in]{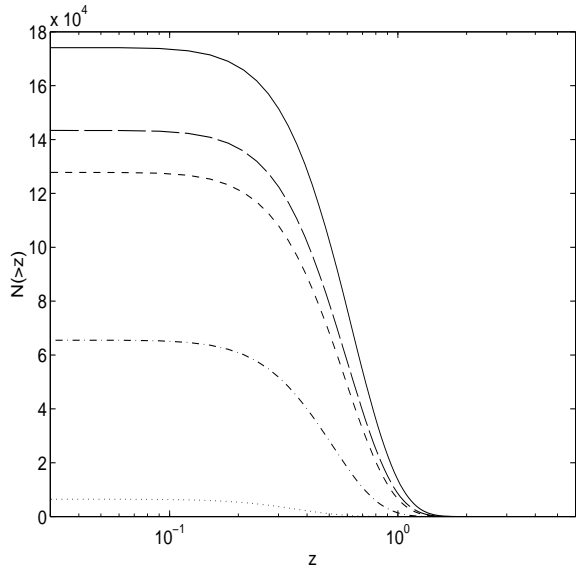}
\caption{Cumulative number counts in redshift space as a function of
  $\sigma_8$. The curves correspond to the same $\sigma_8$ values as in
  fig. 4.}
\end{figure}

In fact, while it is relatively easy to run a code with various input 
data and examine the resulting output, what is of practical interest 
is to follow the reverse course, which is obviously much more difficult 
to implement. Unfortunately, and in contrast with the power spectrum 
of the primary anisotropy, the shape of the S-Z power spectrum is 
rather featureless, implying a high degree of model and parameter 
degeneracy. This means that there are probably several parameter and 
model combinations that result in the same shape for the power spectrum. 
Performing a best likelihood test would necessitate a large collection 
of parameters characterizing both the background cosmology (e.g. 
$\Omega_m$, $\Omega_{\Lambda}$, $h$), the large scale structure 
($\sigma_8$), and cluster properties (\eg IC gas morphology, temperature, 
evolution). The situation will improve significantly when high quality,  
resolved S-Z and X-ray measurements of individual clusters are 
available, along with more precisely determined cosmological 
parameters deduced from independent CMB experiments and large scale 
structure surveys. Various aspects of the task of removing parameter 
degeneracies in the analysis of S-Z and X-ray databases have been 
explored. Mei \& Bartlett (2003) have studied the possibility of 
removing the notorious $\Omega_m$ - $\sigma_8$ degeneracy by combining 
the angular correlation function of S-Z clusters with the observed 
local abundance of X-ray clusters. Diego \& Majumdar (2004) have 
shown how the observed S-Z power spectrum and number counts can be 
combined to construct a ``hybrid power spectrum'' whose advantage 
over the conventional power spectrum is its milder sensitivity to 
$\sigma_8$ and the basic cosmological parameters, thereby increasing 
its diagnostic power of intrinsic cluster properties. Needless to 
say, there is still much room for further theoretical study of S-Z 
power spectrum and number counts in order to devise optimal methods 
of analysis of the wealth of high quality S-Z measurements expected in 
the near future.

\section{Measurements}

In the first two decades following the discovery of the S-Z effect 
measurements were made with single dish ground-based telescopes. 
These observations (which were reviewed in Birkinshaw 1999) are 
plagued by the need to account for the fluctuating atmospheric 
emission. Observational and modeling uncertainties are particularly 
bothersome when observations are made at a single frequency. 
Multi-frequency measurements are essential for separating out the 
various contributing signals, including atmospheric emission, emission 
from Galactic dust, cluster radio sources, and CMB anisotropy. 

Most observational work during the last decade has been made with 
interferometric arrays, whose major advantages over single dish 
telescopes are increased sensitivity to specific angular scales and 
to signals which are correlated between array elements, insensitivity 
to changes in atmospheric emission, and high angular resolution on 
small scales (which is needed to subtract signals from discrete radio 
sources). The improved sensitivity of radio receivers made it feasible 
- mainly through the use of low-noise HEMT amplifiers - to image the 
effect in moderately distant clusters, first with the Ryle telescope 
(Jones \ea 1993), and then mostly with the BIMA and OVRO arrays 
(Carlstrom \ea 1996, 2001). Extensive observations (reviewed by 
Carlstrom \ea 2002) with the latter two arrays, operating at 
frequencies $\sim 30$ GHz, yielded high S/N images of some 60 
clusters in the redshift range $0.17 < z <0.89$, one of which is 
shown in fig. 6.

\begin{figure}
\cl{\includegraphics[height=3in,width=3in]{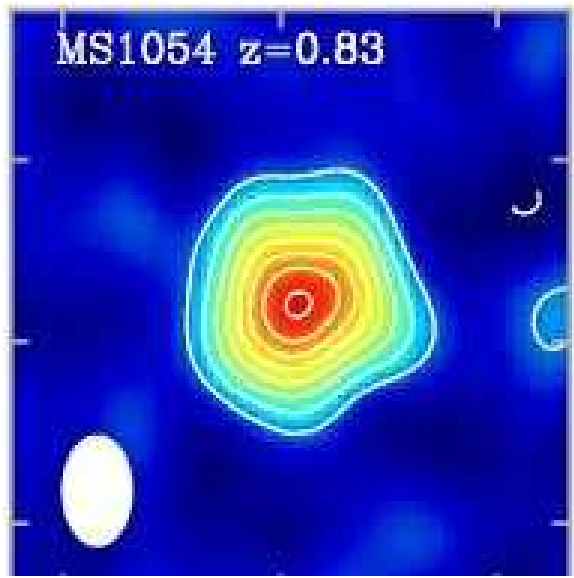}}
\caption{S-Z image of the cluster MS1054 obtained by observations 
at 28.5 GHz with the interferometric BIMA array (from Carlstrom \ea 
2002). Colors mark the magnitude of the (negative) intensity change, 
which is highest (red) in the cluster center.}
\end{figure}  

Interferometric S-Z observations were made also with the CBI, an array
of 13 small (0.9 m) dishes, with $3'-10'$ resolution, operating in the 
$26-36$ GHz spectral range. In contrast to the BIMA and OVRO arrays, 
the compact configuration of the CBI makes this imager suitable for 
observations of nearby clusters. Results from measurements of the 
effect in 7 nearby clusters were reported by Udomprasert \ea (2004).

Multi-frequency measurements of the effect (on the R-J and Wien sides)
are, of course, highly desirable for both the extraction of additional
information from its spectral shape, and for separating the signal out 
from the various sources of confusion. The SuZIE experiment is the first 
multi-element array operating at high frequencies. Some fifteen 
clusters were observed at three frequency bands (centered on $150$, 
$220$, and $350$ GHz) with the (two versions of the) experiment 
(Holzapfel \ea 1997a, 1997b, Mauskopf \ea 2000, Benson \ea 2004). 
High frequency measurements were also made with the balloon-borne 
PRONAOS (Lamarre \ea 1998) - the first detection of the S-Z effect from 
the stratosphere - and ground-based MITO (De Petris \ea 2002) 
experiments. Finally, high spatial resolution measurements of the 
effect in the very luminous cluster RX J1347­1145 were made with the 
NOBA and SCUBA bolometer arrays (Kitayama \ea 2004).

\section{Results from S-Z \& X-ray Measurements}

The main virtues of the S-Z effect that make it a uniquely important
cosmological probe are its physically well understood origin and its 
(essentially) redshift independence. High sensitivity measurements of 
individual clusters directly yield the integrated pressure of hot IC 
gas, and thereby also the total mass of the cluster. In general, IC 
gas density and temperature profiles outside the cluster central region 
can be more directly obtained from spatially resolved S-Z measurements 
of $\Delta I_{t}(r)$ than from measurements of the (more steeply 
falling) X-ray surface brightness profile. When S-Z measurements are 
made close to the crossover frequency, the cluster velocity along 
the los can also be deduced. 

The angular diameter distance, and therefore $H_0$, can be determined 
from S-Z and X-ray measurements. This SZX method to measure $H_0$ has 
clear advantages over the traditional galactic distance ladder method; 
these include its well understood physical nature, and the capability 
to test that the global expansion is indeed isotropic. It is also 
possible to determine the principal contributions to $\Omega$ - 
$\Omega_m$, and the currently favored, $\Omega_{\Lambda}$ - from the 
Hubble diagram and from the mean value of the gas mass fraction 
measured in a (sufficiently large) sample of clusters. 

The feasibility of detecting clusters at large redshifts strongly 
motivates carrying out number counts through cluster surveys in 
order to characterize the population and its cosmological evolution. 
The power spectrum of the CMB anisotropy induced by clusters can 
yield important information on the cluster mass function, cluster 
properties, and the evolution of clusters. Finally, the anisotropy 
and redshift evolution of the CMB temperature, $T(z)$, can be determined 
from multi-frequency measurements of the effect in clusters at 
different sky directions and redshifts.

In the following subsections we review some of the recent results 
obtained from current S-Z measurements. The discussion here is {\it 
very brief} and devoid of any considerations of important observational 
issues, such as impact of confusing signals and related systematic 
uncertainties. Observational aspects are expertly reviewed and assessed
by Birkinshaw in this volume. Comprehensive discussions of results 
from observational S-Z work can be found in the reviews by Birkinshaw 
(1999) and Carlstrom \ea (2002).

\ms
\ni
{\bf The gas mass fraction $f_g$:} The los integrated gas density can 
be directly determined if the projected temperature profile is known 
from spatially resolved spectral X-ray measurements. For a relaxed 
cluster the measured gas density and temperature can be used in the 
hydrostatic equation to deduce the total cluster mass $M(r)$ interior 
to a radial position $r$. The ratio of the gas to total cluster mass, 
$f_g$, is a good measure of the baryonic mass fraction of the cluster, 
$f_b$. Determining this ratio for a sufficiently large region, such 
as $r_{500}$ - the radius at which the mean mass density of the 
cluster is $500$ times the background value - (not only well samples 
this fraction in the cluster but also) provides a fair estimate of 
the universal value of the baryonic density parameter (\eg Evrard 
1997) when averaged over a sufficiently large sample of clusters. 
Current best S-Z estimate for $f_g$ is based on BIMA and OVRO 
measurements of 18 intermediate distance ($0.14<z<0.83$) clusters, 
which yield $f_g h\simeq (0.08 \pm 0.01)h^{-1}$ for the currently 
favored matter, $\Omega_m =0.3$, and cosmological constant, 
$\Omega_{\Lambda}=0.7$, density parameters (Grego \ea 2001). The 
dependence of the gas fraction on the cluster angular diameter 
distance can be used to determine $\Omega_m$. Analysis of the 
above dataset yielded the estimate $\Omega_m \sim 0.25$ in a flat 
$\Lambda$ dominated model with $h=0.7$.

\ms
\ni
{\bf $H_0$ and $\Omega$:} The Hubble constant can be determined from 
measurement of the angular diameter distance of a cluster, $d_A$, 
from resolved S-Z and X-ray observations. When the observed clusters 
span a wide redshift range, the main contributions to the density 
parameter - $\Omega_{m}$ and $\Omega_{\Lambda}$ - can then be deduced 
from the Hubble diagram (a plot of $d_{A}(z)$). Analysis of a database 
of 38 measured distances to 26 clusters (from single dish and 
interferometric BIMA \& OVRO measurements) at redshifts $z\leq 0.83$ 
yielded a mean value of $H_0 = 60 \pm 3$ $km s^{-1}$ Mpc$^{-1}$ 
for a flat model with $\Omega_M = 0.3$ and $\Omega_{\Lambda} = 0.7$ 
(Carlstrom \ea 2002). This $\sim 5\%$ ($1\sigma$) observational error 
is much smaller than the estimated systematic uncertainty of 
$\sim 30\%$. Contributions to the latter include the unknown gas 
thermal profile, the assumptions of sphericity of the gas spatial 
configuration, possible small scale clumping, confusion due to 
CMB primary anisotropy, and the (unknown) cluster peculiar motion 
(see Birkinshaw 1999 for details). 

Clearly, in order for this method for the measurement of $H_0$ to 
yield a value whose level of precision is comparable to those obtained 
by the local distance ladder, the CMB anisotropy, and SN Ia methods, 
the overall level of uncertainty has to be lowered to $\sim 5\%$. 
This could be attained when high quality resolved measurements of a 
large sample of clusters are made. In order to achieve optimal level 
of overall systematic uncertainty, the sample must include many 
{\it nearby} clusters.

\ms
\ni
{\bf Cluster velocities:} Measurements of cluster radial velocities (in 
the CMB frame) from the kinematic S-Z component necessitate sensitive 
observations in a narrow spectral band near the crossover frequency 
(where the thermal component vanishes). The exact spectral shape of 
the thermal component near this frequency needs to be known. Confusing 
signals due to the CMB primary anisotropy may very well be the main 
limitation in reaching levels of precision of a few hundred km/s. 
Attempts to measure radial velocities using this method were 
made mostly with the SuZIE experiment (Holzapfel \ea 1997b, Mauskopf 
\ea 2000, Benson \ea 2003). The SuZIE radial velocity sample includes 
8 clusters with velocities in a wide range, with uncertainties so 
large that none of the deduced values are statistically significant.

\ms
\ni
{\bf CMB temperature:} Measurements with the COBE/FIRAS experiment have 
shown that the CMB spectrum is a precise Planckian with $T_{0} = 
2.725 \pm 0.002 \ $K at the current epoch (Mather \ea 1999). In the 
standard cosmological model, $T(z) = T_{0} (1+z)$, a fundamental 
relation which has not yet been fully confirmed observationally. 
Cosmological models with a purely blackbody spectrum but with a 
different $T(z)$ dependence than in the standard model are - formally, 
at least - unconstrained by the FIRAS measurements. Also unconstrained 
are models with spectral distortions that are now negligible, but may 
have been appreciable in the past. Thus far $T(z)$ has been determined 
mainly from measurements of microwave transitions in interstellar 
clouds in which atoms and molecules are excited by the CMB (\eg LoSecco 
\ea 2001). The temperature was determined in the Galaxy, as well as 
in clouds at redshifts up to $z \sim 3$. Results are, however, 
substantially uncertain due to the poorly known physical conditions 
in the absorbing clouds.

The use of the thermal S-Z effect to measure $T(z)$ was suggested 
long ago (Fabbri, Melchiorri \& Natale 1978, Rephaeli 1980). The 
method proposed by Rephaeli is based on the steep frequency 
dependence of $\Delta I_t$ on the Wien side, and the weak 
dependence of {\it ratios} of the intensity change at different 
frequencies on the properties of the cluster. Formally, in the 
\nrel limit such a ratio is completely independent of the 
Comptonization parameter. Most of the dependence on the cluster 
parameters drops out also in the exact \rel description, but a 
weak dependence remains on the gas temperature; the (unknown) 
cluster velocity introduces a small systematic uncertainty. 
S-Z measurements have the potential of yielding more precise 
values of $T(z)$ than can be obtained from ratios of atomic and 
molecular lines. 

The availability of spectral measurements of the S-Z effect enabled 
implementation of the method of Rephaeli (1980) to measure $T(z)$ 
in the Coma and A2163 clusters; the measurements and their analysis 
are described by Battistelli \ea (2002), and by Melchiorri \& 
Olivo-Melchiorri in this volume. Spectral measurements of Coma 
($z= 0.0231 \pm 0.0017$) and A2163 ($z=0.203 \pm 0.002$) at 
four frequency bands yield three independent intensity ratios 
for each cluster; all combinations of these ratios were compared 
to the theoretically predicted values. Fits of the measured ratios 
to the predicted values were performed, yielding best fit values for 
the CMB temperature at the redshifts of the two clusters, $T_{Coma} 
= 2.789^{+0.080}_{-0.065}$ K and $T_{A2163} = 3.377^{+0.101}_{-0.102}$ 
K (at 68\% confidence). These values are consistent with those 
expected from the standard relation $T(z)=T_{0}(1+z)$. Battistelli 
\ea (2002) have also tested two alternative scaling relations that 
are conjectured in non-standard cosmologies, $T(z)= T_{0}(1+z)^{1-a}$, 
and  $T(z)=T_{0}[1+(1+d)z]$ (e.g., Lima et al. 2000). They determined 
the best fit values for the two parameters to be 
$a=-0.16^{+0.34}_{-0.32}$, and $d = 0.17 \pm 0.36$ (at 95\% confidence), 
values that are consistent with zero, so no significant deviation was 
found from the standard model. LoSecco \ea (2001) obtained 
$a=-0.05\pm 0.13$ and $d= 0.10\pm 0.28$ (at 95\% CL) from measurements 
of microwave transitions. The two sets of results are consistent. 
Thus, the S-Z results of Battistelli \ea (2002) already provide the 
same level of precision even though the two clusters are at much lower 
redshifts than the galaxies in the sample used by LoSecco \ea (2001). 
With more precise spectral S-Z measurements expected in the future, 
it is anticipated that the S-Z method will provide a preferred 
alternative to the atomic and molecular lines method.

\section{Prospects for the Near Future}

The quality of the scientific yield from the many S-Z images obtained
with the interferometric BIMA and OVRO arrays proved beyond doubt 
that the S-Z effect is an indispensable cosmological probe whose great 
potential has just begun to be exploited. New projects have either 
already begun observations of the S-Z effect or will become operational 
in the next few years. These include the ground based AMI, ACT, AMiBA, 
APEX, MAD/MITO, and SPT, the stratospheric project OLIMPO, and the 
Planck satellite, whose full sky survey is expected to result in the 
detection of thousands of clusters (as reviewed by Hansen in this 
volume).

With the much improved multi-frequency and high spatial resolution 
capabilities of essentially all the new S-Z projects, the effect 
will be mapped in many hundreds of clusters, in both pointed and 
survey modes, thereby greatly enhancing the scope of the measurements 
and the quality of the scientific yield. Deep, high resolution 
measurements of the effect in many nearby clusters will provide 
the best database for precise characterization of the S-Z properties 
of clusters. The detailed spectral and spatial images will not 
only yield state-of-the-art gas and total mass profiles, but will 
also provide us with the key knowledge to assess and reduce the 
overall level of systematic uncertainties. This will lead to a 
great improvement in the precision of the derived values of cluster 
masses and of the Hubble constant. Extensive cluster surveys and 
the measurement of the S-Z induced CMB anisotropy will yield the 
mass and redshift distributions of clusters. Important information 
will be extracted on the evolution of clusters and the large scale 
structure. Cosmological parameters, including the dark energy 
content of the universe, will be determined at a high level of 
precision.


\begin{thebibliography}{99}
\frenchspacing

\bibitem{} Battistelli, B.S., \ea 2002, \apj, 580, L101
\bibitem{} Birkinshaw M. 1999, Phys. Rep., 310, 97
\bibitem{} Bartlett J., \& Silk J., 1994, \apj, 423, 12
\bibitem{} Benson, B.A. 2003, \apj, 592, 674
\bibitem{} Benson, B.A. 2004, \apj, 617, 829
\bibitem{} Bond J.R., Ruetalo M.I., Wadsley J.W., \& Gladders M.D.,2002, 
{\it ASPC}, 257, 15
\bibitem{} Carlstrom, J.E., Joy, M., \& Grego, L. 1996, ApJ, 456, L75
\bibitem{} Carlstrom,J.E., Holder, G.P., \& Reese, E.D.\ 2002, ARA\& A, 40, 643
\bibitem{} Challinor, A., \& Lasenby, A.\ 1998, \apj, 499, 1
\bibitem{} Challinor, A.D., Ford, M.T., \& Lasenby, A.N.\ 2000, MNRAS, 312, 159
\bibitem{} Chandrasekhar, S. 1950, `Radiative Transfer', Oxford, Clarendon Press
\bibitem{} Colafrancesco S., Mazzotta P., Rephaeli Y., \& Vittorio N. 1994, 
\apj, 433, 454
\bibitem{} Colafrancesco S., Mazzotta P., Rephaeli Y., \& Vittorio N., 
1997 ,\apj, 479, 1
\bibitem{} Cooray A., Hu W., \& Tegmark M., 2000, \apj, 540, 1
\bibitem{} Cooray, A., \& Baumann, D.\ 2003, \prd, 67, 063505 
\bibitem{} da Silva A.C., Barbosa D., Liddle A.R., \& Thomas P.A. 2000, 
\mn, 317, 37
\bibitem{} De Petris, M., \ea 2002, ApJ, 574, L119
\bibitem{} Diego, J.~M., Mazzotta, P., \& Silk, J. 2003, \apjl, 597, L1 
\bibitem{} Diego J.M., \& Majumdar S., 2004, \mn, 352, 993
\bibitem{} Evrard, A. 1997, \mn, 292, 289
\bibitem{} Fabbri, R., Melchiorri, F., \& Natale, V. 1978, ApS\&S, 59, 223
\bibitem{} Grego, L., \ea 2001, ApJ, 552, 2
\bibitem{} Holzapfel, W.L., \ea 1997a, ApJ, 480, 449
\bibitem{} Holzapfel, W.L., \ea 1997b, ApJ, 481, 35
\bibitem{} Hu, W.\ 2000, \apj, 529, 12 
\bibitem{} Hughes, J.P., \& Birkinshaw, M., 1998, \apj, 501, 1
\bibitem{} Itoh, N., Nozawa, S., \& Kohyama, Y. 2000, ApJ, 533, 588
\bibitem{} Itoh, N., \& Nozawa, S. 2004, A\&A, 417, 827 
\bibitem{} Jones, M., et al. 1993, Nature, 365, 320
\bibitem{} Kamionkowski, M., \& Loeb, A. 1997, PRD, 56, 4511
\bibitem{} Kitayama, T., \ea 2004, Pub. Astron. Soc. Japan, 56, 17
\bibitem{} Komatsu E., \& Kitayama T. 1999, \apj, 526, 1
\bibitem{} Komatsu E., \& Seljak U. 2002, MNRAS, 336, 1256
\bibitem{} Lamarre, J.M., \ea 1998, ApJ, 507, L5
\bibitem{} Lavaux, G., Diego, J.M., Mathis, H., \& Silk, J. 2004, 
MNRAS, 347, 729
\bibitem{} Lee J., \& Shandarin S.F. 1999, \apj, 517, 5
\bibitem{} Lima, J.A.S., \ea 2000, MNRAS, 312, 747
\bibitem{} LoSecco, J.M., \& 2001, Phys. Rev. D 64, 123002
\bibitem{} Makino N., \& Suto Y. 1993, \apj, 405, 1
\bibitem{} Mather, J.C., \ea 1995, \apj, 512, 511
\bibitem{} Mauskopf, P., \ea 2000, ApJ, 538, 535
\bibitem{} Mei S., \& Bartlett J.G. 2003, \aa, 410, 767
\bibitem{} Molnar S.M., \& Birkinshaw M. 2000, \apj, 537, 542
\bibitem{} Nozawa, S., Itoh, N., \& Kohyama, Y. 1998a, \apj, 502, 7
\bibitem{} Nozawa, S., Itoh, N., \& Kohyama, Y.\ 1998b, \apj, 508, 17
\bibitem{} Ohno, H., Takada, M., Dolag, K., Bartelmann, M., \& Sugiyama, N. 
2003, \apj, 584, 599 
\bibitem{} Press W.H., \& Schechter P. 1974, \apj, 187, 425
\bibitem{} Rephaeli Y. 1980, \apj, 241, 858
\bibitem{} Rephaeli Y. 1981, \apj, 245, 351
\bibitem{} Rephaeli Y. 1995a, ARA\& A, 33, 541
\bibitem{} Rephaeli Y. 1995b, \apj, 445, 33
\bibitem{} Rephaeli, Y., \& Yankovitch, D. 1997, \apjl, 481, L55 
\bibitem{} Sadeh,S., \& Rephaeli, Y. 2004, New Astronomy, 9, 159
\bibitem{} Sazonov, S.Y., \& Sunyaev, R.~A. 1998, \apj, 508, 1
\bibitem{} Sazonov, S.Y., \& Sunyaev, R.~A. 1999, MNRAS, 310, 765
\bibitem{} Sheth R.K., \& Tormen G. 1999, \mn, 308, 119
\bibitem{} Shimon, M., \& Rephaeli, Y. 2004, New Astronomy, 9, 69
\bibitem{} Shimon M., Rephaeli Y., O'Shea B.W., \& Norman M.~L. 2005, 
submitted 
\bibitem{} Springel V., White M., \& Hernquist L. 2001, \apj, 549, 681
\bibitem{} Sunyaev R. 1977, Comm. Ap. Sp. Phys., 7, 1
\bibitem{} Sunyaev, R.A., \& Zeldovich, Y.B. 1972, Comm. Ap. Sp. 
Phys., 4, 173
\bibitem{} Sunyaev, R.A., \& Zeldovich, Y.B. 1980, MNRAS, 190, 413
\bibitem{} Udomprasert, P.S., \ea 2004, \apj 615, 63
\bibitem{} Weymann R. 1966, \apj 145, 560
\bibitem{} Wright E.L. 1979, \apj, 232, 348-351
\bibitem{} Zeldovich, Y.B., \& Sunyaev, R.A. 1969, Astrophys. Sp. 
Sci., 4, 30l
\bibitem{} Zhang P., Pen, U.L., \& Wang B. 2002, \apj, 577, 555

\end{thebibliography}
\end{document}